    \documentclass[twocolumn,final]{svjour3}
\smartqed  
\usepackage{graphicx}
 \usepackage{mathptmx}      
    \usepackage{bm}

    \usepackage{dcolumn}

\usepackage{amsmath}
\usepackage{txfonts}

    \usepackage{amssymb}

    \usepackage{showkeys}

    \usepackage[usenames]{color}

\def\bal#1\eal{\begin{align}#1\end{align}}
\newcommand\beq{\begin{equation}}
\newcommand\eeq{\end{equation}}
\newcommand\beqa{\begin{eqnarray}}
\newcommand\eeqa{\end{eqnarray}}
\newcommand{\nn}{\nonumber\\}

\newcommand{\dd}{{\text{d}}}
\newcommand{\ee}{{\text{e}}}

\newcommand{\ancho}{6cm}
\newcommand{\anchob}{5cm}

\newcommand{\ma}{m_i}
\newcommand{\mb}{m_j}
\newcommand{\cca}{\mathbf{v}_i}
\newcommand{\ca}{\varv_i}

\newcommand{\ccb}{\mathbf{v}_j}

\newcommand{\cc}{\mathbf{v}}

\newcommand{\kk}{\widehat{\bm{\sigma}}}
\newcommand{\wwa}{\bm{\omega}_i}
\newcommand{\wwb}{\bm{\omega}_j}
\newcommand{\wa}{{\omega}_i}

\newcommand{\ww}{\bm{\omega}}
\newcommand{\Ia}{I_i}

\newcommand{\da}{\sigma_i}
\newcommand{\db}{\sigma_j}
\newcommand{\ds}{\sigma}
\newcommand{\dab}{\sigma_{ij}}
\newcommand{\x}{\times}

\newcommand{\een}{\alpha_{ij}}
\newcommand{\esn}{\alpha}
\newcommand{\eet}{\beta_{ij}}
\newcommand{\est}{\beta}

\newcommand{\enn}{\overline{\alpha}_{ij}}

\newcommand{\ett}{\overline{\beta}_{ij}}
\newcommand{\mab}{m_{ij}}
\newcommand{\qab}{\kappa_{ij}}
\newcommand{\qa}{\kappa_{i}}
\newcommand{\qb}{\kappa_{j}}
\newcommand{\q}{\kappa}

\newcommand{\Tat}{T_{i}^\text{tr}}
\newcommand{\Tbt}{T_{j}^\text{tr}}
\newcommand{\Tt}{T^\text{tr}}
\newcommand{\Tar}{T_{i}^\text{rot}}
\newcommand{\Tbr}{T_{j}^\text{rot}}
\newcommand{\Tr}{T^\text{rot}}

\newcommand{\zabt}{\xi_{ij}^\text{tr}}
\newcommand{\zt}{\xi^\text{tr}}
\newcommand{\zabr}{\xi_{ij}^\text{rot}}
\newcommand{\zr}{\xi^\text{rot}}
\newcommand{\al}{i}
\newcommand{\be}{j}
\newcommand{\tr}{\text{tr}}
\newcommand{\rot}{\text{rot}}

\newcommand{\dt}{d_{\text{tr}}}
\newcommand{\dr}{d_{\text{rot}}}
\newcommand{\Fwn}{\mathbf{F}^{\text{wn}}}
\newcommand{\wn}{\text{wn}}
\newcommand{\Xo}{\chi_0}

 \journalname{Granular Matter}
\begin{document}

\title{Driven and undriven states of multicomponent granular gases of inelastic and rough hard disks or spheres
}

                        \titlerunning{Driven and undriven states of multicomponent granular gases}        

\author{Alberto Meg\'ias \and Andr\'es Santos
}


\institute{A. Meg\'ias \and A. Santos \at
              Departamento de F\'{\i}sica, Universidad
de Extremadura, E-06006 Badajoz, Spain \\
                           }

\date{\today}

\maketitle

\begin{abstract}
Starting from a recent derivation  of the energy production rates in terms of the number of translational  and rotational  degrees of freedom,
a comparative study on  different granular temperatures  in  gas mixtures of inelastic and rough disks or spheres is carried out. Both the homogeneous freely cooling state and the state driven by a  stochastic thermostat are considered. It is found that the relaxation number of collisions per particle  is generally smaller for disks than for spheres, the  mean angular velocity relaxing more rapidly than the temperature ratios. In the asymptotic regime of the undriven system, the rotational-translational nonequipartition is stronger in disks than in spheres, while it is hardly dependent on the class of particles in the driven system. On the other hand, the degree of component-component nonequipartition is higher for spheres than for disks, both for driven and undriven systems. A study of the mimicry effect (whereby  a multicomponent gas mimics the rotational-translational temperature ratio of a monocomponent gas) is also undertaken.

\end{abstract}

\section{Introduction}

This paper is dedicated to the memory of  Robert P.\ Behringer, who paved the way for a better understanding of dense granular matter. His long lasting influence and impact on the field can be appreciated in part from an excellent (posthumous) review paper \cite{BC19}.

While the basic model of a granular gas is a collection of inelastic and \emph{smooth}  hard disks or spheres, either monodisperse
\cite{
BP04,D00,G03} or polydisperse
\cite{
BT02a,DHGD02,G19,GD99b,GD02,GDH07,GM07,JM89,MG02b,SGNT06,UKAZ09}, the model can be significantly
improved by incorporating the rotational degrees of freedom of the particles (assumed to be \emph{rough})
\cite{
BPKZ07,CP08,G19,GSK18,GNB05,HZ97,JR85a,KSG14,L95,LHMZ98,LB94,LS87,MHN02,MSS04,SKS11,SP17,VLSG17b,VS15,VSK14,ZTPSH98}.
The aim of this paper is to employ kinetic-theory methods to compare the degrees of breakdown of energy equipartition in hard disks and spheres when both roughness and polydispersity are considered. Due to the  angular motion inherent to roughness, the distinction between   disks  and spheres is not trivial.
In contrast to spinning disks on a plane, which have $\dt=2$ translational and $\dr=1$ rotational degrees of freedom, spinning spheres in space have $\dt=3$ translational plus $\dr=3$ rotational degrees of freedom.

In a recent work \cite{MS18}, we have presented a \emph{unified} kinetic-theory derivation (in terms of the number of degrees of freedom $\dt$ and $\dr$) of
the collisional rates of energy production in multicomponent granular gases, so that
previous results for disks \cite{S18} and spheres \cite{SKG10} are obtained by taking $(\dt,\dr)=(2,1)$ and $(\dt,\dr)=(3,3)$, respectively. Those unified expressions will be applied here to study the granular temperature ratios in monodisperse and bidisperse gases of rough disks or spheres in homogeneous states, both undriven and driven.
Experimental realizations of those states can be found, for instance, in Refs.\ \cite{GBG09,HTWS18,MIMA08,TMHS09} and \cite{CR91,FM02,GSVP11b,HKTSHWS13,HTMWS15,HYCMW04,PGGSV12,TMHS09,WP02,YHCMW02} for the undriven and driven cases, respectively.

The remainder of the paper is organized as follows. Section \ref{sec2} defines the systems and presents the energy production rates. This is followed by the application to the homogeneous cooling state (HCS) and to the state driven by a stochastic thermostat in Sects.\ \ref{sec3} and \ref{sec4}, respectively. Section \ref{sec5} deals with the conditions for a mixture to mimic a monocomponent gas in what concerns the temperature ratios (mimicry effect). Finally, the conclusions are presented in Sect.\ \ref{sec6}.

\section{Energy production rates}
\label{sec2}
Let us consider a dilute multicomponent granular gas made of hard disks or spheres of different masses $\{\ma\}$, diameters $\{\da\}$, and moments of inertia $\{\Ia\}$. We denote by $\cca$ and $\wwa$ the translational and angular velocities, respectively, of a particle belonging to component $i$. In a binary collision between particles of components $i$ and $j$, linear total momentum is conserved, as is the angular momentum of each particle with respect to the point of contact, but this is not enough to determine the postcollisional velocities in terms of the precollisional ones and the unit vector $\kk$ pointing from the center of particle $i$ to the center of particle $j$.
To close the collision rules, it is frequently assumed that the normal and tangential components of the relative velocity
$\mathbf{w}_{ij}\equiv\cca-\ccb-\kk\x(\da\wwa+\db\wwb)/2$ of the points of the particles at contact become, after collision,
$\mathbf{w}_{ij}'\cdot\kk=-\een\mathbf{w}_{ij}\cdot\kk$, $\kk\x\mathbf{w}_{ij}'=-\eet\kk\x\mathbf{w}_{ij}$,
where $\een$ and $\eet$ are the coefficients of normal and tangential restitution, respectively, assumed here to be constant. The coefficient $\een$ ranges from $\een=0$ (perfectly inelastic particles) to $\een=1$ (perfectly elastic particles). In contrast, $\eet$ ranges from  $\eet=-1$ (perfectly smooth particles) to $\eet=1$ (perfectly rough particles). It can be easily checked that the total (translational plus rotational) kinetic energy is a collisional invariant only if $\een=|\eet|=1$.

The mean values of the translational and rotational kinetic energies of particles of component $i$ define the so-called (partial) \emph{granular} temperatures, namely \cite{SKG10} $T_i^\tr={m}\langle \ca^2\rangle/{\dt}$, $T_i^\rot={I}_i\langle \wa^2\rangle/{\dr}$,
where, as said before, $\dt$ and $\dr$ are the number of translational and rotational degrees of freedom, respectively, and a zero mean translational velocity has been assumed. Analogously, one can define the mean angular velocity of component $i$ as $\bm{\Omega}_i=\langle \wwa\rangle$.
The rates of change of the quantities $\bm{\Omega}_{i}$, $\Tat$, and $\Tar$ due to collisions with particles of component $j$ can be written as
\begin{subequations}
\bal
\left.\partial_t\bm{\Omega}_{i}\right|_{\text{coll},j}=&-\frac{\zeta_{ij}^\Omega}{2\da}\left(\da\bm{\Omega}_i+{\sigma_j}\bm{\Omega}_j\right),\\
\left.\partial_t\Tat\right|_{\text{coll},j}=&-\zabt\Tat,\quad\left.\partial_t\Tar\right|_{\text{coll},j}=-\zabr\Tar,
\eal
\end{subequations}
where $\zeta_{ij}^\Omega$ are spin production rates, and $\zabt$ and $\zabr$ are energy production rates. While the exact determination of those quantities is not possible, a kinetic-theory approach (namely the Boltzmann equation) supplemented by a multitemperature Maxwellian approximation \cite{MS18,S18,SKG10} allows one to express them in terms of the partial densities ($n_i$, $n_j$), temperatures ($\Tat$, $\Tar$, $\Tbt$, $\Tbr$), mean angular velocities ($\bm{\Omega}_{i}$, $\bm{\Omega}_{j}$), and the mechanical parameters. The unified expressions for disks ($\dt=2$, $\dr=1$) and spheres ($\dt=\dr=3$) are \cite{MS18}
\begin{subequations}
\label{4abc}
\bal
\zeta_{ij}^{\Omega}=&\frac{\nu_{ij}}{\dt}\frac{4\mab\ett}{\ma\qa},
\quad \nu_{ij}\equiv\frac{\sqrt{2}\pi^{\frac{\dt-1}{2}}}{\Gamma(\dt/2)}n_j\sigma_{ij}^{\dt-1}\sqrt{\frac{\Tt_i}{m_i}+\frac{\Tt_j}{m_j}},\\
\label{4b}
\zabt=&\frac{\nu_{ij}}{\dt}\frac{ 2\mab^2 }{\ma \Tat }\Bigg[2\left(\enn+\frac{\dr}{\dt}\ett\right)\frac{\Tat}{\mab}\nn
&-\left(\enn^2+\frac{\dr}{\dt}\ett^2\right)\left(\frac{\Tat}{\ma}+\frac{\Tbt}{\mb}\right)\nn
&-\frac{\dr}{\dt}\ett^2\left(\frac{\Tar}{\ma\qa}+\frac{\Tbr}{\mb\qb}+\frac{\da\db\bm{\Omega}_{\al}\cdot\bm{\Omega}_{\be}}{2\dr} \right)\Bigg],\\
\label{4c}
\zabr=&\frac{\nu_{ij}}{\dt}\frac{4\mab^2\ett}{\ma\qa\Tar}\Bigg[\frac{\Tar}{\mab}+\frac{\ma\qa}{\mab}
\frac{\da\db\bm{\Omega}_{\al}\cdot\bm{\Omega}_{\be}}{4\dr}\nn
&
-\frac{\ett}{2}\left(\frac{\Tat}{\ma}+\frac{\Tbt}{\mb}+\frac{\Tar}{\ma\qa}+\frac{\Tbr}{\mb\qb}+
\frac{\da\db\bm{\Omega}_{\al}\cdot\bm{\Omega}_{\be}}{2\dr}\right)\Bigg].
\eal
\end{subequations}
Here, $\qa\equiv\Ia/(\ma\da^2/4)$ is a reduced moment of inertia, $\mab\equiv {\ma\mb}/{(\ma+\mb)}$ is the reduced mass, $\sigma_{ij}\equiv\frac{1}{2}(\da+\db)$, $\enn\equiv 1+\een$, and $\ett\equiv (1+\eet){\qab}/{(1+\qab)}$, where
 $\qab\equiv \qa\qb{(\ma+\mb)}/{(\qa\ma+\qb\mb)}$.
Equation \eqref{4b} generalizes to the rough case results previously derived for smooth spheres \cite{BLB14,GD02,GM07,UKAZ09}.
In the special case of a monocomponent gas, Eqs.\ \eqref{4abc} become
\begin{subequations}
\label{6abc}
  \bal
  \zeta^\Omega=&\frac{2\nu}{\dt}\frac{1+\beta}{1+\kappa},\quad \nu\equiv\frac{2\pi^{\frac{\dt-1}{2}}}{\Gamma(\dt/2)} n\sigma^{\dt-1}\sqrt{\frac{\Tt}{m}},\\
   \xi^\tr=&\frac{\nu}{\dt}\Big\{1-\alpha^2+\frac{2\dr\kappa(1+\beta)}{\dt(1+\kappa)^2 T^\tr}\Big[\frac{\kappa(1-\beta)}{2}\Big(T^\tr
  +\frac{T^\rot}{\kappa}\nn
&+\frac{m\sigma^2\Omega^2}{4\dr}\Big)
+T^\tr-T^\rot-\frac{{\kappa}m\sigma^2\Omega^2}{4\dr}\Big]\Big\},\\
\xi^\rot=&\frac{2\nu}{\dt}\frac{\kappa(1+\beta)}{(1+\kappa)^2T^\rot}\Big[
\frac{1-\beta}{2}\Big(T^\tr+\frac{T^\rot}{\kappa}\nn
&+\frac{m\sigma^2\Omega^2}{4\dr}\Big)
+T^\rot-T^\tr+\frac{{\kappa}m\sigma^2\Omega^2}{4\dr}\Big].
\eal
\end{subequations}

It is interesting to remark that, in the case of  smooth inelastic spheres, a similar multitemperature Maxwellian approximation has been used for isotropic mixtures \cite{GD02} and for monocomponent granular fluids with horizontal-vertical anisotropy \cite{vMR06}.
Apart from that type of Maxwellian approximation for the distribution of translational velocities, the derivation of Eqs.\ \eqref{4abc} and \eqref{6abc}
is based on the assumption that the statistical correlations between translational and angular velocities can be neglected. On the other hand,
couplings between $\cc$ and $\ww$ have been predicted theoretically and confirmed by simulations \cite{BPKZ07,KBPZ09,SKS11,VS15,VSK14} in the case of spheres ($\dt=\dr=3$). Nevertheless, those couplings are relatively small; for instance, at $\alpha=0.9$, one has $|\langle \varv^2\omega^2\rangle/ \langle \varv^2\rangle\langle\omega^2\rangle-1|<0.08$ and $|\langle (\cc\cdot \ww)^2\rangle/\langle \varv^2\omega^2\rangle-\frac{1}{3}|<0.05$ both in the driven \cite{VS15} and undriven \cite{BPKZ07,KBPZ09,VSK14} systems.

\section{Undriven gas: homogeneous cooling state}
\label{sec3}
In the HCS, time evolution of the mean values is due to collisions only. In particular,
$\partial_t\Tat=-\zt_i\Tat$, $\partial_t\Tar=-\zr_i\Tar$,
where $\zt_i\equiv\sum_j\zabt$ and $\zr_i\equiv\sum_j\zabr$. After a certain transient period, the gas reaches a long-time asymptotic regime where all temperatures decay with a common rate \cite{BLB14,S11b,VLSG17,VSK14}, so that the temperature ratios are obtained from the conditions $\zt_1=\zt_2=\cdots=\zr_1=\zr_2=\cdots$. Our goal now is to compare those ratios in the cases of hard disks and hard spheres. To that end, we will suppose a uniform mass distribution in both types of particles, so that the reduced moment of inertia is $\qa=\frac{1}{2}$ for disks and $\qa=\frac{2}{5}$ for spheres.

\subsection{Monocomponent system}
Given the three energy scales $\Tt$, $\Tr$, and $I\Omega^2$, we can construct the following two dimensionless quantities: $\theta\equiv \Tr/\Tt$ and $X\equiv I\Omega^2/\dr\Tr=\langle\bm{\omega}\rangle\cdot \langle\bm{\omega}\rangle/\langle\omega^2\rangle$.
Using Eqs.\ \eqref{6abc}, one can obtain in a straightforward way a coupled set of nonlinear differential equations for the evolution of $\theta$ and $X$:
\beq
\label{evol_HCS}
\frac{1}{2}\partial_\tau \ln \theta+\xi^{\rot,*}-\xi^{\tr,*}=\frac{1}{2}\partial_\tau \ln X+2\zeta^{\Omega,*}-\xi^{\rot,*}=0,
\eeq
where a star denotes division by the collision frequency $\nu$ (i.e., $\xi^{\tr,*}\equiv \zt/\nu$, etc.) and $\tau=\frac{1}{2}\int_0^t\dd t'\, \nu(t')$ is the accumulated number of collisions per particle.
Taking into account that $X<1$, it can be easily checked that $2\zeta^{\Omega,*}-\xi^{\rot,*}$ is positive definite, thus implying that
$\lim_{\tau\to\infty}X(\tau)=0$. On the other hand, the evolution equation for $\theta(\tau)$ admits a stationary solution,
$\theta_s$, given by the condition $\xi^{\tr,*}=\xi^{\rot,*}$, which yields the quadratic equation
$\theta_s-1-\frac{\dt}{\dr}\left(\theta_s^{-1}-1\right)=2h$,
whose physical solution is
\begin{subequations}
\label{theta_inf}
\bal
\theta_{s}=&\sqrt{\left[h+ \frac{1}{2}\left(1-\frac{\dt}{\dr}\right)\right]^2+\frac{\dt}{\dr}}+h+\frac{1}{2}\left(1-\frac{\dt}{\dr}\right),\\
h\equiv&\frac{\dt(1+\kappa)^2}{2\dr\kappa(1+\beta)^2}\left[1-\alpha^2-\frac{1-\frac{\dr}{\dt}\kappa}{1+\kappa}(1-\beta^2) \right].
\eal
\end{subequations}

\begin{figure}
\begin{center}
\includegraphics[width=\anchob]{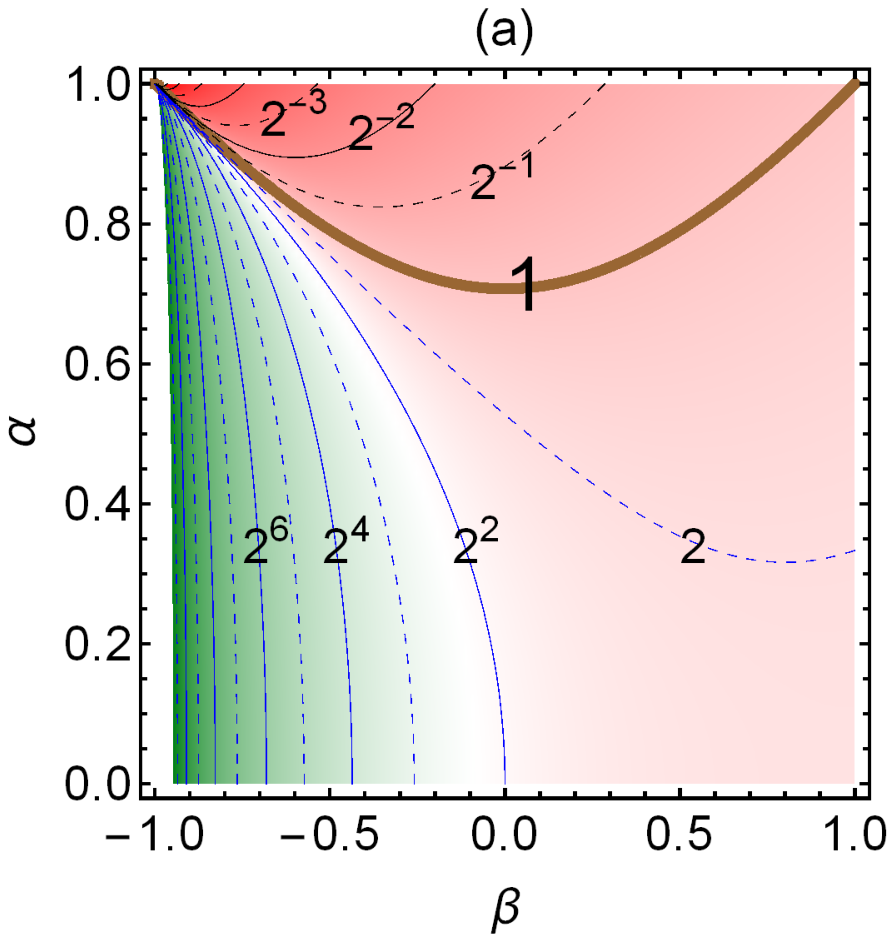}\\
\includegraphics[width=\anchob]{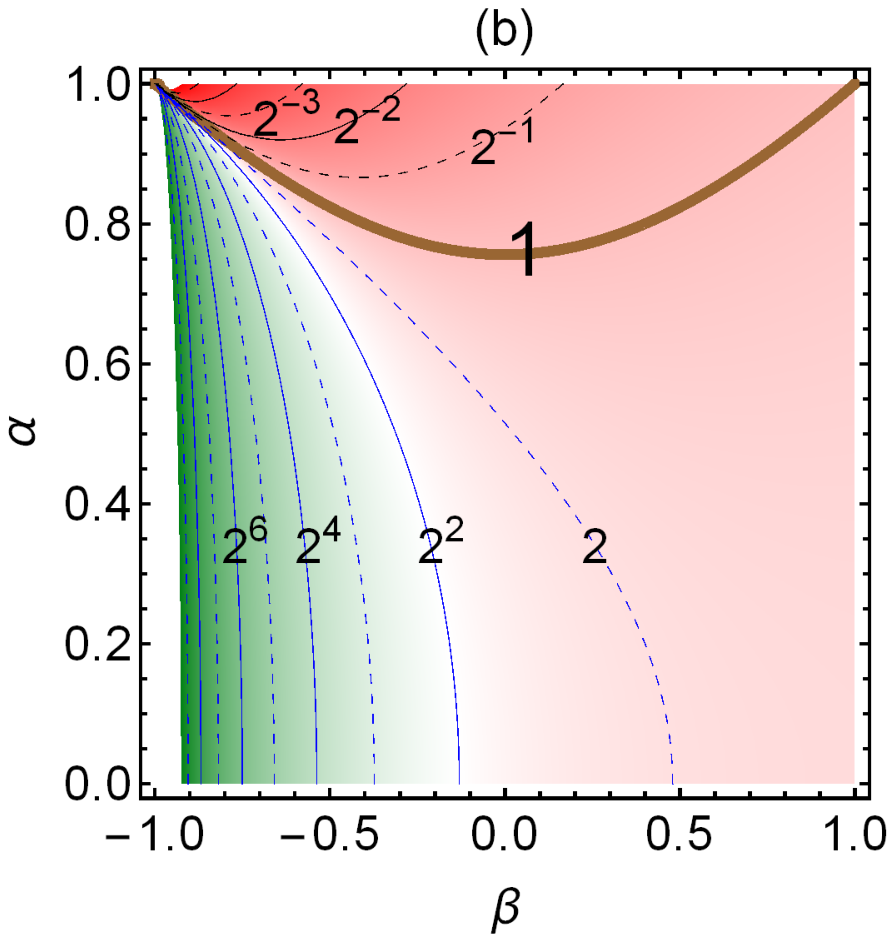}
\end{center}
\caption{Density plot of  the stationary value $\theta_s$ of the temperature ratio $\theta\equiv \Tr/\Tt$ in the HCS [see Eqs.\ \eqref{theta_inf}] for (a) uniform disks ($\q=\frac{1}{2}$) and (b) uniform spheres ($\q=\frac{2}{5}$). The contour lines correspond to $\theta_s=1$ (thick solid line), $\theta_s=2^{-1}, 2^{-2}, 2^{-3}, \ldots$, and $\theta_s=2, 2^{2}, 2^{3}, \ldots$.
\label{fig1}}
\end{figure}

Figure \ref{fig1} shows a density plot of the stationary temperature ratio $\theta_s$ as a function of the coefficients of restitution $\alpha$ and $\beta$ in the cases of (a) uniform disks ($\q=\frac{1}{2}$) and (b) uniform spheres ($\q=\frac{2}{5}$). In both cases, the equipartition line $\theta_s=1$ (where $h=0$)  splits the plane $(\beta,\alpha)$ into two regions. In the lower region, the rotational temperature is higher than the translational one ($\theta_s>1$, $h>0$), while the opposite occurs in the upper region.
Apart from those common features, we can observe that, in general, the breakdown of rotational-translational equipartition is higher in disks than in spheres.

\begin{figure}
\begin{center}
\includegraphics[width=\ancho]{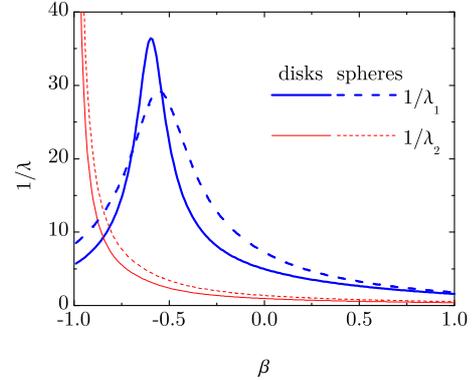}
\end{center}
\caption{Plot of the relaxation times $1/\lambda_1$ (thick blue lines) and $1/\lambda_2$ (thin red lines) versus $\beta$ at $\alpha=0.8$  in the HCS [see Eqs.\ \eqref{eigenvalues}] for  uniform disks (solid lines) and uniform spheres (dashed lines).
\label{fig2}}
\end{figure}

Once the stationary solution $(\theta,X)=(\theta_s,0)$ of the HCS is established, it is convenient to analyze its stability. Linearization of the evolution equations \eqref{evol_HCS} yields the solution
\begin{subequations}
\bal
\delta\theta(\tau)=&\delta\theta_0\ee^{-\lambda_1\tau}-\frac{\lambda_{12}}{\lambda_2-\lambda_1}X_0
\left(\ee^{-\lambda_1\tau}-\ee^{-\lambda_2\tau}\right),\\
X(\tau)=&X_0\ee^{-\lambda_2\tau},
\eal
\end{subequations}
where $\delta\theta(\tau)\equiv\theta(\tau)-\theta_s$ and
\begin{subequations}
\label{eigenvalues}
\bal
\lambda_1=&\frac{2\kappa}{\dt}\left(\frac{1+\beta}{1+\kappa}\right)^2\left(\frac{1}{\theta_s}+\frac{\dr}{\dt}\theta_s\right),\\
\lambda_2=&\frac{2}{\dt}\frac{1+\beta}{1+\kappa}\left[2+(1+\beta)\frac{1+\kappa/\theta_s}{1+\kappa}\right],\\
\lambda_{12}=&\frac{2\theta_s}{\dt}\frac{1+\beta}{(1+\kappa)^2}\left[2\kappa+1-\beta+\frac{\dr}{\dt}(1+\beta)\kappa\theta_s\right].
\eal
\end{subequations}
As expected on physical grounds, both eigenvalues $\lambda_1$ and $\lambda_2$ are positive definite, what confirms the linear stability of the stationary solution $(\theta,X)=(\theta_s,0)$ with respect to homogeneous perturbations. The quantities $1/\lambda_1$ and $1/\lambda_2$ are the characteristic relaxation times (measured as the number of collisions per particle) associated with the evolution of $\delta\theta(\tau)$ (if $X_0=0$ or $\lambda_1<\lambda_2$) and $X(\tau)$, respectively. Both relaxation times are plotted in Fig.\ \ref{fig2} as functions of the coefficient of tangential restitution at the representative value $\alpha=0.8$. It can be observed that, except for very small roughness ($-1\leq\beta\lesssim 0.84$), one has $1/\lambda_2<1/\lambda_1$. This justifies the non-hydrodynamic character of the angular velocity in a hydrodynamic description \cite{KSG14}.
As for the difference between disks and spheres, Fig.\ \ref{fig2} also shows that, in general, disks require a smaller number of collisions than spheres to reach the stationary state. The only exception is the interval $-0.69\lesssim\beta\lesssim -0.54$, where $1/\lambda_1$ is larger for disks than for spheres.

\subsection{Binary system}

\begin{figure}
\begin{center}
\includegraphics[width=\ancho]{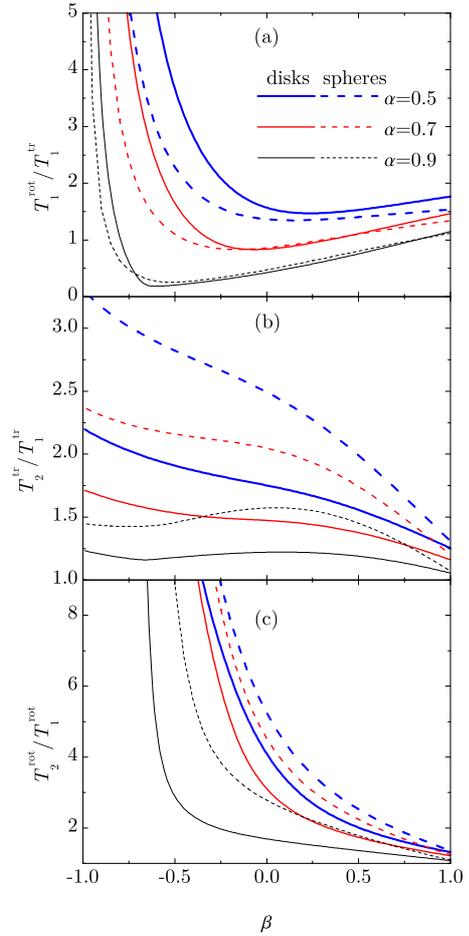}
\end{center}
\caption{Plot of the temperature ratios (a) $\Tr_1/\Tt_1$, (b) $\Tt_2/\Tt_1$, and (c) $\Tr_2/\Tr_1$  versus $\beta$ in the HCS for  equimolar binary mixtures of uniform disks (solid lines) or uniform spheres (dashed lines) with $\sigma_2/\sigma_1=2$, $m_2/m_1=2^{\dt}$,  $\beta_{ij}=\beta$, and
$\alpha_{ij}=\alpha=0.5$ (thick blue lines), $0.7$ (medium red lines),  and $0.9$ (thin black lines).
\label{fig3}}
\end{figure}

As a representative multicomponent gas, let us consider here a binary system that has already reached the asymptotic HCS. The conditions $\xi_1^\tr=\xi_1^\rot=\xi_2^\tr=\xi_2^\rot$ provide the three independent temperature ratios ($\Tr_1/\Tt_1$, $\Tt_2/\Tt_1$, and $\Tr_2/\Tr_1$) for arbitrary values of the $11$ dimensionless parameters of the system ($n_2/n_1$, $m_2/m_1$, $\sigma_2/\sigma_1$, $\q_1$, $\q_2$, $\alpha_{11}$, $\alpha_{12}$, $\alpha_{22}$,  $\beta_{11}$, $\beta_{12}$, and $\beta_{22}$). For the sake of concreteness, we will consider an equimolar mixture ($n_2/n_1=1$) where all the particles are uniform ($\q_i=\frac{1}{2}$ and $\frac{2}{5}$ for disks and spheres, respectively) and  made of the same material (i.e., $\alpha_{ij}=\alpha$ and  $\beta_{ij}=\beta$). Moreover, the size of the large particles is assumed to be twice that of the small particles ($\ds_2/\ds_1=2$), so that $m_2/m_1=2^{\dt}$.

Figure \ref{fig3} shows the three independent temperature ratios as functions of the roughness parameter $\est$ for a few characteristic values of the inelasticity parameter $\esn$.
The rotational-translational temperature ratio $\Tr_1/\Tt_1$ has a behavior qualitatively similar to
that of the monodisperse case (see Fig.\ \ref{fig1}) in the sense that $\Tr_1/\Tt_1<1$ if $\alpha$ is larger than a certain threshold value and $\beta$ belongs to a certain $\alpha$-dependent interval around $\beta\approx 0$, whereas $\Tr_1/\Tt_1>1$ otherwise. Also, the departure from rotational-translational equipartition ($\Tr_1/\Tt_1=1$) is generally stronger for disks than for spheres. In contrast, Figs.\ \ref{fig3}(b) and \ref{fig3}(c) show that the translational and rotational component-component ratios exhibit a stronger nonequipartition effect in the case of spheres than in the case of disks.

\section{Driven gas: Stochastic thermostat}
\label{sec4}

Let us consider now a homogeneous dilute granular gas subject to a stochastic volume force $\Fwn$ (also called a thermostat), which injects translational kinetic energy to the particles and has the properties of a Gaussian white noise \cite{BT02a,MS00,vNE98,WM96}, i.e.,
$\langle \Fwn_{\al} (t)\rangle=\mathbf{0}$,
$\langle \Fwn_{\al}(t) \Fwn_{\be} (t')\rangle = \mathsf{I}m_i^2\Xo^2\delta_{\al\be}\delta(t-t')$,
where indices $\al$, $\be$ refer to particles, $\mathsf{I}$ is the $\dt \times \dt$ unit matrix, and $\Xo^2$ measures the strength of the stochastic force.
This kind of forcing can model, for example, the energy input to grains immersed in a gas in turbulent flow.

Since the stochastic force acts on the translational degrees of freedom only, the time evolutions of the mean angular velocities ($\bm{\Omega}_i$) and the rotational temperatures ($\Tar$) are governed by collisions only. On the other hand, $\partial_t\Tat=\ma \Xo^2-\zt_i \Tat$.
The conditions for a stationary state are $\bm{\Omega}_i=\mathbf{0}$, $\zr_i=0$, and $\ma \Xo^2=\zt_i \Tat$.

\subsection{Monocomponent system}
Apart from $X\equiv I\Omega^2/\dr \Tr$ and $\theta\equiv \Tr/\Tt$, the stochastic thermostat introduces a third dimensionless parameter, $Y\equiv m \Xo^2/\nu\Tt$, which can be seen as a (time-dependent) reduced measure of the noise strength. Instead of Eq.\ \eqref{evol_HCS}, now we have
\bal
\label{evol_wn}
\frac{1}{2}\partial_\tau \ln \theta+\xi^{\rot,*}-\xi^{\tr,*}+Y=&
\frac{1}{2}\partial_\tau \ln X+2\zeta^{\Omega,*}-\xi^{\rot,*}\nn
=&\frac{1}{2}\partial_\tau \ln Y+\frac{3}{2}\left(Y-\xi^{\tr,*}\right)=0.
\eal
As in the undriven case,  $2\zeta^{\Omega,*}-\xi^{\rot,*}$ is positive definite, so that $\lim_{\tau\to\infty}X(\tau)=0$. Moreover, $\xi^{\rot,*}=0$ and $Y-\xi^{\tr,*}=0$ give  the stationary values
\begin{equation}
\theta_{s}^\wn=\frac{1+\beta}{2+\kappa^{-1}(1-\beta)},
\quad Y_s=\frac{1-\alpha^2}{\dt}+\frac{2\dr}{\dt^2}(1-\beta)\theta_{s}^\wn.
\end{equation}
Note that $\theta_{s}^\wn$ is independent of $\alpha$, $\dt$, and $\dr$. However,  it depends on the reduced moment of inertia $\q$, so that it is slightly larger for uniform disks ($\q=\frac{1}{2}$) than for uniform spheres ($\q=\frac{2}{5}$).  Since $\theta_{s}^\wn\leq 1$, this implies that, in contrast to the HCS case,  the degree of rotational-translational nonequipartition is higher in spheres than in disks.

\begin{figure}
\begin{center}
\includegraphics[width=\ancho]{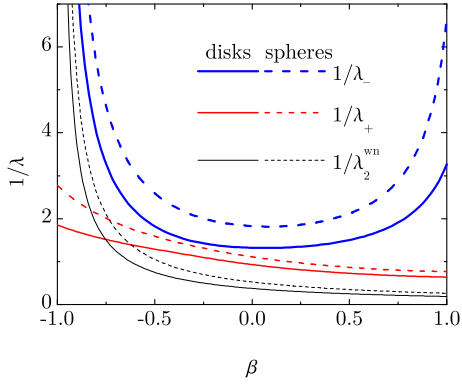}
\end{center}
\caption{Plot of the relaxation times $1/\lambda_-$ (thick blue lines), $1/\lambda_+$ (medium red lines), and $1/\lambda_2^\wn$ (thin black lines) versus $\beta$ at $\alpha=0.8$  in systems driven by a stochastic thermostat [see Eqs.\ \eqref{eigenvalues_wn}] for  uniform disks (solid lines) and uniform spheres (dashed lines).
\label{fig4}}
\end{figure}

As in the HCS case, it is instructive to analyze the time evolution of $\delta\theta\equiv \theta-\theta_s^{\wn}$, $\delta Y\equiv Y-Y_s$ and $X$ near the stationary state. After linearizing Eqs.\ \eqref{evol_wn}, one obtains
\begin{subequations}
\bal
\delta\theta(\tau)+\frac{\lambda_\pm-\lambda_{1}^\wn}{\lambda_{31}}\delta Y(\tau)=&\left(\delta\theta_0+\frac{\lambda_\pm-\lambda_{1}^\wn}{\lambda_{31}}\delta Y_0\right)\ee^{-\lambda_\pm\tau}\nn
&-\frac{\lambda_{\pm}\theta_{s}^\wn X_0}{\lambda_{2}^\wn-\lambda_\pm}
\left(\ee^{-\lambda_\pm\tau}-\ee^{-\lambda_{2}^\wn\tau}\right),\\
X(\tau)=&X_0\ee^{-\lambda_{2}^\wn\tau},
\eal
\end{subequations}
where
\begin{subequations}
\label{eigenvalues_wn}
\bal
\lambda_\pm=&\frac{1}{2}\left[\lambda_{1}^\wn+3Y_s\pm\sqrt{(\lambda_{1}^\wn-3Y_s)^2+8\theta_{s}^\wn\lambda_{31}}\right],\\
\lambda_{1}^\wn=&\frac{2\kappa}{\dt}\left(\frac{1+\beta}{1+\kappa}\right)^2\left(\frac{1}{\theta_{s}^\wn}+\frac{\dr}{\dt}\theta_{s}^\wn\right),\\
\lambda_{2}^\wn=&\frac{8}{\dt}\frac{1+\beta}{1+\q},\quad \lambda_{31}=\frac{3\dr\q}{\dt^2}\left(\frac{1+\beta}{1+\q}\right)^2Y_s.
\eal
\end{subequations}
The dependence on $\beta$ of the reciprocal  eigenvalues $1/\lambda_\pm$ and $1/\lambda_2^\wn$ is shown in Fig.\ \ref{fig4} at $\alpha=0.8$.
Comparison with Fig.\ \ref{fig2} shows that the relaxation toward the stationary values is much faster in the driven gas than in the undriven one. Although $1/\lambda_+<1/\lambda_2^\wn$ if $\beta\lesssim -0.74$, one has $1/\lambda_->1/\lambda_2^\wn$ for all $\beta$, so that the reduced angular velocity $X$ tends to zero much more rapidly than $\delta\theta$ and $\delta Y$. Finally, in agreement with the undriven case, we can observe that the relaxation times (as measured by the number of collisions per particle) are shorter for disks than for spheres.

\subsection{Binary system}

\begin{figure}
\begin{center}
\includegraphics[width=\ancho]{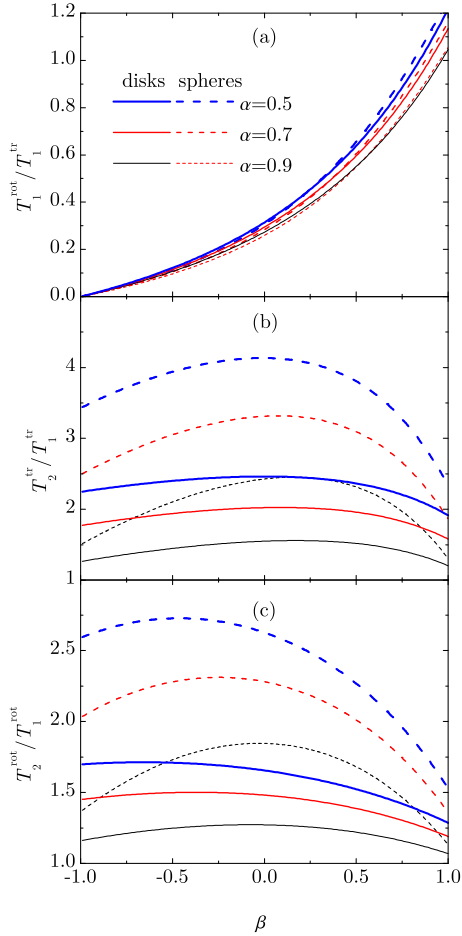}
\end{center}
\caption{Plot of the temperature ratios (a) $\Tr_1/\Tt_1$, (b) $\Tt_2/\Tt_1$, and (c) $\Tr_2/\Tr_1$  versus $\beta$ in systems driven by a stochastic thermostat for  equimolar binary mixtures of uniform disks (solid lines) or uniform spheres (dashed lines) with $\sigma_2/\sigma_1=2$, $m_2/m_1=2^{\dt}$,  $\beta_{ij}=\beta$, and $\alpha_{ij}=\alpha=0.5$ (thick blue lines), $0.7$ (medium red lines),  and $0.9$ (thin black lines).
\label{fig5}}
\end{figure}

In the case of a binary mixture driven by a stochastic thermostat, the three independent temperature ratios ($\Tr_1/\Tt_1$, $\Tt_2/\Tt_1$, and $\Tr_2/\Tr_1$) in the steady state are obtained by the conditions $\zr_1=\zr_2=0$ and $\zt_1\Tt_1/m_1=\zt_2\Tt_2/m_2$. Again, we choose here an equimolar mixture ($n_2/n_1=1$) with $\q_i=\frac{1}{2}$ and $\frac{2}{5}$ for disks and spheres, respectively, $\alpha_{ij}=\alpha$,  $\beta_{ij}=\beta$, $\ds_2/\ds_1=2$, and $m_2/m_1=2^{\dt}$.

The temperature ratios are shown in Fig.\ \ref{fig5}  as functions of the roughness parameter $\est$ for the same values of $\esn$ as in Fig.\ \ref{fig3}.
The rotational-translational temperature ratio $\Tr_1/\Tt_1$ exhibits a very weak dependence on $\alpha$ and is hardly sensitive to whether the particles are disks or spheres. While in the monocomponent case the degree of rotational-translational nonequipartition is slightly higher in spheres than in disks, from Fig.\ \ref{fig5}(a) one can observe that this ceases to be true for large enough roughness in the case of mixtures. As for the translational and rotational component-component ratios, the equipartition breakdown is clearly stronger for spheres than for disks, in analogy to what happens in the undriven case (see Fig.\ \ref{fig3}).

\section{Mimicry effect}
\label{sec5}
As illustrated  by Figs.\ \ref{fig3} and \ref{fig5},  in the long-time asymptotic regime each component of a mixture has in general a different translational ($\Tt_1\neq\Tt_2\neq\cdots$) and rotational ($\Tr_1\neq\Tr_2\neq\cdots$) temperature, both in the driven and the undriven states, even if all the coefficients of restitution and all the reduced moments of inertia are equal ($\een=\alpha$, $\eet=\beta$, $\qa=\qab=\q$).
In general, if the particle mass densities are similar, the bigger particles have larger temperatures. This is exemplified in a high-component mixture of smooth spheres ($\dt=3$, $\eet=-1$) with $m_i\propto \sigma_i^3\propto i$ and a sufficiently steep size composition $n_i$; in that case, the temperature of the bigger spheres follows the scaling law $\Tat\propto \ma^\gamma$, with $\gamma\simeq 1.85$ and $1.22$ for the undriven and driven systems, respectively \cite{BLB14}.

In this context, an interesting question \cite{S18} is whether it is possible to couple the densities, sizes, and masses of the particles in such a way that all the components reach a common translational temperature ($\Tt_i=\Tt$) and a common rotational temperature ($\Tr_i=\Tr$). In that case, the temperature ratio $\Tr_i/\Tt_i=\Tr/\Tt$ would be the same as that of a monocomponent gas and one can say that the mixture \emph{mimics} the monocomponent system.

\begin{figure}
\begin{center}
\includegraphics[width=\ancho]{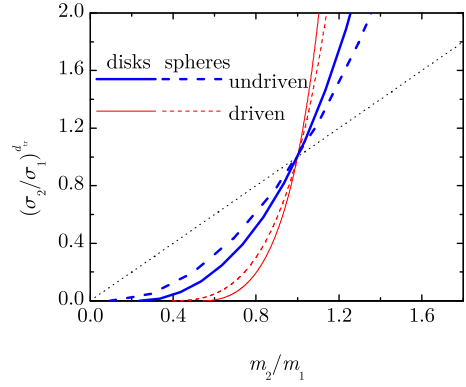}
\end{center}
\caption{Plot of the area or volume ratio $(\sigma_2/\sigma_1)^{\dt}$  versus the mass ratio $m_2/m_1$ for the mimicry effect in  equimolar binary mixtures of disks (solid lines) or spheres (dashed lines). The thick blue and thin red curves correspond to the undriven and driven systems, respectively. The dotted straight line represents the points $(\sigma_2/\sigma_1)^{\dt}=m_2/m_1$ of  equal particle mass density.
\label{fig6}}
\end{figure}

Setting $\een=\alpha$, $\eet=\beta$, $\qa=\q$, $\Tt_i=\Tt$,  $\Tr_i=\Tr$, and $\bm{\Omega}_i=\mathbf{0}$ in Eqs.\ \eqref{4b} and \eqref{4c}, we obtain
\beq
\label{20bis}
\xi_i^{\tr,\rot}=\xi_{11}^{\tr,\rot} R_i, \quad R_i\equiv\sum_j\frac{n_j\dab^{\dt-1}}{n_1\sigma_1^{\dt-1}}\sqrt{\frac{2m_1m_j}{m_i(m_i+m_j)}}.
\eeq
According to Eq.\ \eqref{20bis}, the HCS conditions $\xi^\tr_1=\xi^\tr_2=\cdots=\xi^\rot_1=\xi^\rot_2=\cdots$ decompose into $\xi^\tr_{11}=\xi^\rot_{11}$ (which actually is the monocomponent condition) plus $R_1=R_2=\cdots$ (which establish constraints on densities, diameters, and masses for the mimicry effect).
In the driven case, Eq.\ \eqref{20bis} shows that $\xi_i^\rot=0$ implies  $\xi_{11}^\rot=0$ (monocomponent condition), while $\xi_1^\tr T_1^\tr/m_1=\xi_2^\tr T_2^\tr/m_2=\cdots$ imply $R_1/m_1=R_2/m_2=\cdots$.
It is remarkable that those mimicry conditions are independent of the coefficients of restitution ($\alpha$ and $\beta$), the reduced moment of inertia ($\kappa$), and, in the case of the driven system, the noise strength ($\Xo^2$).

To fix ideas, let us consider a binary mixture. The conditions $R_1=R_2$ (undriven system) and $R_1/m_1=R_2/m_2$ (driven system) yield
\begin{equation}
\label{22}
\frac{n_2}{n_1}=\frac{\sigma_{12}^{\dt-1}\left(\frac{m_1}{m_2}\right)^\epsilon\sqrt{\frac{m_1}{m_2}}- \sigma_1^{\dt-1} \left(\frac{m_2}{m_1}\right)^\epsilon \sqrt{\frac{m_1+m_2}{2m_1}}}{\sigma_{12}^{\dt-1}\left(\frac{m_2}{m_1}\right)^\epsilon\sqrt{\frac{m_2}{m_1}}- \sigma_{2}^{\dt-1}\left(\frac{m_1}{m_2}\right)^\epsilon\sqrt{\frac{m_1+m_2}{2m_2}}},
\end{equation}
where $\epsilon=0$ and $\frac{1}{2}$ in the undriven and driven cases, respectively.
Equation \eqref{22}  represents the  constraint on $n_2/n_1$, $\sigma_2/\sigma_1$, and $m_2/m_1$ for the mimicry effect. By solving a linear equation in the case of disks ($\dt=2$) or a quadratic equation in the case of spheres ($\dt=3$), it is possible to express $\sigma_2/\sigma_1$ as explicit functions of $n_2/n_1$ and $m_2/m_1$. If $m_2/m_1\approx 1$, one has $\sigma_2/\sigma_1-1=k(m_2/m_1-1)/(\dt-1)$ with independence of the density ratio, where $k=\frac{3}{2}$ and $\frac{7}{2}$ for undriven and driven gases, respectively. It is interesting to notice that the mass ratio $m_2/m_1$ must be larger than a lower bound (corresponding to $\sigma_2/\sigma_1\to 0$) and smaller than an upper bound (corresponding to $\sigma_2/\sigma_1\to \infty$). More specifically,
$\mu_0(n_2/n_1)< {m_2}/{m_1}<{1}/{\mu_0(n_1/n_2)}$,
where
\beq
\label{23b}
\mu_0(x)=\frac{x+2^{\dt-2}\left[2^{\dt-2}-\sqrt{2^{2(\dt-2)}+2(1+x)}\right]}{x^2-2^{2\dt-3}}
\eeq
for undriven gases, while $\mu_0(x)$ is the positive real root of the quartic equation $2^{2\dt-3}\mu_0^3(1+\mu_0)=(1-x\mu_0^2)^2$ for driven gases.

Figure \ref{fig6} shows the area (in the case of disks) or volume (in the case of spheres) ratio $(\sigma_2/\sigma_1)^{\dt}$ as a function of the mass ratio $m_2/m_1$, as obtained from Eq.\ \eqref{22} in the equimolar case ($n_2/n_1=1$). If $m_2<m_1$, one has $(\sigma_2/\sigma_1)^{\dt}<m_2/m_1$, i.e., $m_2/\sigma_2^{\dt}>m_1/\sigma_1^{\dt}$, while the opposite happens if $m_2>m_1$. Therefore, the mimicry effect requires that the smaller particles have a higher particle mass density than the large spheres, this property holding for any $n_2/n_1$. The disparity in the particle mass density is stronger for disks than for spheres and in driven than in undriven systems. In fact, for equimolar mixtures, the windows of mass ratios are $0.094<m_2/m_1<10.657$, $0.236<m_2/m_1<4.236$, $0.398<m_2/m_1<2.510$, and $0.544<m_2/m_1<1.839$ for undriven spheres, undriven disks, driven spheres, and driven disks, respectively.
It is worth mentioning that a recent work \cite{LVGS19} shows a good agreement between theory and computer simulations for the mimicry effect in undriven hard spheres.

\section{Conclusions}
\label{sec6}
In this paper we have carried out a comparative study on the partition of the mean kinetic energy among different classes of degrees of freedom in multicomponent granular gases of disks or spheres. Both undriven (HCS) and driven (stochastic thermostat) states have been considered. The starting point has been a recent unified derivation (within a Maxwellian approximation) of the energy production rates \cite{MS18} in terms of the number of translational ($\dt$) and rotational ($\dr$) degrees of freedom.

The main conclusions are the following ones: (i) the number of collisions per particle needed to reach stationary values for the temperature ratios is generally smaller for disks than for spheres and in the driven system than in the undriven one; (ii) except in the HCS near the quasi-smooth limit, the relaxation time for the mean angular velocity is much shorter than for the temperature ratios; (iii) while in the driven case the rotational-translational temperature ratio is very similar for disks and spheres, in the undriven case disks typically present a stronger rotational-translational nonequipartition than spheres; (iv) on the other hand, the degree of component-component nonequipartition is higher for spheres than for disks, both for driven and undriven systems; (v) under certain conditions, a multicomponent gas can mimic a monocomponent gas in what concerns the rotational-translational temperature ratio; (vi) this mimicry effect requires the smaller component to have a higher particle mass density than the larger component, this property being more pronounced in the driven system than in the undriven one and for disks than for spheres; (vii) interestingly, a mixture mimicking a monocomponent gas in the undriven state loses its mimicry property in the driven steady state (no matter the intensity of the stochastic force), and vice versa.

Before closing this paper, it is worth remarking that our analytic results have been obtained within the framework  of the standard collision
model where both coefficients of restitution are constant. However, more realistic models, with the coefficients of restitution depending on the normal and tangential components of the impact velocity $\mathbf{w}_{ij}$, have been proposed in the literature \cite{BSSP04,BP04,RPBS99,SBHB10,SBP08}. Notwithstanding this,  the experimental measurement of the normal coefficient of restitution at very small impact velocities is challenging \cite{DF17}, some independent experiments \cite{GBG09,SLCL09} providing evidence on a sharp decrease at small impact velocities, in contrast to what happens with viscoelastic spheres \cite{BP04,RPBS99}.

\begin{acknowledgements}
The research of A.S. has been supported by the Agencia Estatal de Investigaci\'on (Spain) through Grant No.\ FIS2016-76359-P and by the Junta de Extremadura
(Spain) through Grant No.\ GR18079, both partially financed by Fondo Europeo de Desarrollo Regional funds.
\end{acknowledgements}

\section*{Compliance with ethical standards}
\paragraph*{Conflict of interest: } The authors declare that they have no conflict of
interest.



\end{document}